\pdfoutput=1
\documentclass[pdf,pdftex,%
preprint %
]{iucr}%

\def\href#1{\relax}\let\foo\caption
\ifPDF
\usepackage{amsmath}
\usepackage{graphicx}
\usepackage{SIunits}
\usepackage[T1]{fontenc}
\usepackage[utf8]{inputenc}
\usepackage[textsize=footnotesize]{todonotes}
\usepackage{hyperref}

\fi
\let\caption\foo

\renewcommand{\harvardurl}[1]{\textbf{URL:} \urlstyle{same} \textit{\url{#1}}}

\paperprodcode{a000000}      % Replace with production code if known
\paperref{xx9999}            % Replace xx9999 with reference code if known
\papertype{FA}               % Indicate type of article
\paperlang{english}          % Can be english, french, german or russian
\journalcode{J}              % Indicate the journal to which submitted
\journalyr{2017}
\journalreceived{\relax}
\journalaccepted{\relax}
\journalonline{\relax}

\begin{document}
	\title{Femtosecond x-ray diffraction from an aerosolized beam of protein nanocrystals}
	\author[a,b]{Salah}{Awel}
	\author[e]{Richard}{A.\ Kirian}
	\author[a,c]{Max}{O.\ Wiedorn}
	\author[a]{Kenneth}{R.\ Beyerlein}
	\author[a]{Nils}{Roth}
	\author[a,b]{Daniel}{A.\ Horke}
	\author[a]{Dominik}{Oberthür}
	\author[a]{Juraj}{Knoska}
	\author[a]{Valerio}{Mariani}
	\author[a]{Andrew}{Morgan}
	\author[d]{Luigi}{Adriano}
	\author[a,c]{Alexandra}{Tolstikova}
	\author[a,h]{P.\ Lourdu}{Xavier}
	\author[a]{Oleksandr}{Yefanov}
	\author[g]{Andrew}{Aquila}
	\author[a]{Anton}{Barty}\
	\author[e]{Shatabdi}{Roy-Chowdhury}
	\author[g]{Mark}{S.\ Hunter}
	\author[e]{Daniel}{James}
	\author[g]{Joseph}{S.\ Robinson}
	\author[e]{Uwe}{Weierstall}
	\author[f]{Andrei}{V.\ Rode}
	\author[d]{Sa\v{s}a}{Bajt}
	\author[a,b,c]{Jochen}{Küpper}
	\cauthor[a,b,c]{Henry}{N.\ Chapman}{henry.chapman@desy.de}
	
	\aff[a]{Center for Free-Electron Laser Science, Deutsches Elektronen-Synchrotron DESY, Notkestraße
		85, 22607 \city{Hamburg}, \country{Germany}}%
	\aff[b]{The Hamburg Center for Ultrafast Imaging, Universität Hamburg, Luruper Chaussee 149, 22761
		\city{Hamburg}, \country{Germany}}%
	\aff[c]{Department of Physics, Universität Hamburg, Luruper Chaussee 149, 22761 \city{Hamburg},
		\country{Germany}}%
	\aff[d]{Photon Science, Deutsches Elektronen-Synchrotron DESY, Notkestraße 85, 22607 Hamburg,
		\country{Germany}}%
	\aff[e]{Arizona State University, \city{Temp}e, Arizona, \country{USA}}%
	\aff[f]{Laser Physics Centre, Research School of Physics and Engineering, Australian National
		University, ACT 2601, \city{Canberra}, \country{Australia}}%
	\aff[g]{Linac Coherent Light Source (LCLS), SLAC National Accelerator Laboratory, 2575, Sand Hill
		Road, \city{Menlo Park}, CA 94025, \country{USA}}%
	\aff[h]{Max-Planck Institute for the Structure and Dynamics of Matter, Luruper Chaussee 149, 22761
		\city{Hamburg}, \country{ Germany}}%
	
	\shortauthor{Awel \emph{et al.}}
	\shorttitle{x-ray diffraction off aerosolized nanocrystals}
	% \keyword{keyword}
	% \PDBref[optional name]{refcode}
	% \NDBref[optional name]{refcode}
	\maketitle
	
	\begin{synopsis}
		A new particle-injection approach is demonstrated that achieves very low background in the
		measurement of diffraction from macromolecular nanocrystals by using an aerosol-focusing injector
		with an x-ray free-electron laser.
	\end{synopsis}
	
	\begin{abstract}
		We demonstrate high-resolution Bragg diffraction from aerosolized single granulovirus
		nanocrystals using an x-ray free-electron laser. The outer dimensions of the in-vacuum aerosol
		injector components are identical to conventional liquid-microjet nozzles used in serial
		diffraction experiments, which allows the injector to be utilized with standard mountings. As
		compared with liquid-jet injection, the x-ray scattering background is reduced by several orders
		of magnitude by the use of helium carrier gas rather than liquid. Such reduction is required for
		diffraction measurements of small macromolecular nanocrystals and single particles. High particle
		speeds are achieved, making the approach suitable for use at upcoming high-repetition-rate
		facilities.
	\end{abstract}

	\section{Introduction}
	\label{sec:introduction}
	Serial femtosecond crystallography (SFX) allows the structural analysis of macromolecular crystals
	that may be too small or weakly scattering to study with synchrotron radiation sources. In order to
	record any measurable diffraction signal, such samples would require exposures far in excess of
	limits imposed by x-ray induced radiation damage when using conventional sources. With typical pulse
	energies of about 1~mJ, or $10^{12}$ photons, and durations of about 10~fs, pulses from x-ray
	free-electron lasers (XFELs) overcome this limit by producing diffraction data before the onset of
	most damage processes~\cite{Neutze:Nature406:752,Chapman:Nature470:73,Boutet:Science337:362}.
	Furthermore, XFELs enable novel time-resolved studies with femtosecond temporal resolution, angstrom spatial resolution, all at physiological temperatures. A variety of prominent results from
	SFX measurements are summarized in recent reviews and special issues~\cite{Spence:RPP75:102601,Schlichting:COSB22:613,Barty:ARPC64:415,Patterson:CR20:242,Schlichting:IUCRJ2:246,Johansson:TBS42:749}. 
	
	The large increase in x-ray fluence afforded by the ability to outrun damage not only increases the
	diffraction signal from the sample, but it also increases the diffuse scattering from the medium
	transporting the crystal to the beam. Many SFX measurements were, therefore, performed on
	microcrystals that were large enough and ordered well-enough to produce intense Bragg reflections
	that could be measured in the presence of the diffuse background. Such examples helped the rapid
	adoption of the technique. % the reviews above are the best citations of adoption
	The use of such crystals, usually with volumes greater than $\unit{1}{\micro\metre^3}$, enabled a
	broad range of sample delivery methods to be used depending on the nature of the experiment, such as
	liquid microjets~\cite{DePonte:JPD41:195505,Weierstall:PTRSB369:20130337}, viscous extrusion
	injectors~\cite{Weierstall:NatComm5:3309} or solid substrates~\cite{Frank:IUCRJ1:95,Roedig:Nm14:805}.
	This flexibility is in stark contrast to efforts to record high-resolution coherent diffraction
	patterns from non-crystalline samples~\cite{Seibert:Nature470:78,Kuepper:PRL112:083002,Aquila:SD2:041701,Yang:PRL117:153002}. Without the amplfication of the diffraction signal due to periodicity,
	objects such as molecules, viruses, and other particles produce only weak scattering signals.
	Non-crystalline samples must, therefore, be delivered to the x-ray focus in a vacuum environment and
	in isolation from other potential scattering sources. This can be achieved, for example, through
	aerodynamic focusing of aerosolised particles \cite{Bogan:NanoLett8:310,Bogan:AST44:i,Bogan:JPB43:194013,Roth:arXiv}. In certain cases, background scattering from a dense surrounding
	medium is highly undesirable even for experiments on crystalline samples. For example, imaging
	techniques have been developed to utilize the faint continuous diffraction signal in regions between
	and at scattering angles beyond the intense Bragg reflections due to lattice
	disorder~\cite{Ayyer:Nature530:202,Chapman:JAC50:1084} or lattice truncation~\cite{Spence:OE19:2866,Kirian:PRX5:011015}. The advantage and motivation for such approaches is that the continuous
	diffraction that can be accessed provides a direct route to solving the crystallographic phase
	problem without the need for prior knowledge or additional measurements.
	
	\sloppy%
	Here, we demonstrate high-resolution x-ray diffraction from isolated protein nanocrystals delivered
	into the XFEL focus via a convergent-nozzle aerosol injector (CNAI)~\cite{Kirian:SD2:041717}. We
	show that the aerosol delivery produces extremely low background scattering signals compared with a
	conventional liquid jet. This aerosol injector has essentially the same size and form as the nozzles
	that are commonly used to produce liquid jets for SFX experiments~\cite{DePonte:JPD41:195505,Beyerlein:RSI86:125104,Oberthuer:SR7:44628} and hence can be installed using standard liquid-jet
	mountings available at x-ray facilities. As shown in our previous work, CNAIs can produce
	aerosolized beams of sub-micrometer particles with a full-width at half maximum diameter
	\text{$<\unit{5}{\micro\metre}$} and particle velocities on the order of a few hundred meters per
	second, depending on particle size and operating conditions. This high velocity may be well-suited
	to the MHz repetition rates of upcoming XFEL sources.	
	
	\section{Experimental methods}
	\label{sec:methods}
	This proof-of-principle experiment was performed on natural Cydia pomonella granulovirus (GV)
	particles of approximately $200\times200\times370~\text{nm}^3$ in size that consist of a central
	virus body surrounded by a crystalline granulin protein shell. They infect invertebrates such as the
	codling moth (Cydia pomonella)~\cite{Jehle:VJ346:180}. The GV particles used in this study were
	purified from a biopesticide solution (Certis Madex HP) using a method described
	elsewhere~\cite{Oberthuer:SR7:44628} and suspended in water at a concentration of approximately
	$3\times10^{11}$ particles/ml prior to injection. The particle concentration was measured using a
	NanoSight (model LM14C) particle analysis system. The volume of the particle is about
	$\unit{0.015}{\micro\metre^3}$, with about 2/3 of that found as the volume of the
	crystalline shell~\cite{Gati:PNAS114:9}, which corresponds to a diameter of approximately 300~nm for
	a sphere of equivalent volume. Despite their small size, previous SFX experiments recorded
	diffraction to 2.1~\AA{} resolution from such nanocrystals delivered to the x-ray beam in a gas-focused
	liquid jet~\cite{Gati:PNAS114:9}.
	
	Diffraction measurements were performed in the nanofocus chamber at the coherent x-ray imaging
	(CXI)~\cite{Liang:JSR22:514} instrument at the Linac Coherent Light Source (LCLS). The experiment
	was carried out immediately after a successful liquid-jet experiment~\cite{Oberthuer:SR7:44628}
	without disruption to the x-ray beam. During that earlier experiment the beam focus was optimised by
	adjusting the Kirkpatrick-Baez (KB) focusing mirrors with the help of analysis of spot imprints on a
	gold foil. After optimisation the position of the beam was determined by placing a YAG screen in the
	focal plane and observing optical fluorescence with a fixed in-line microscope with a resolution of
	a few ${\micro\meter}$. In our experiment, the aerosol beam was initially aligned relative to this reference,
	and then scanned  in position as described below.
	
\begin{figure}
		\centering
		\includegraphics[width=\linewidth]{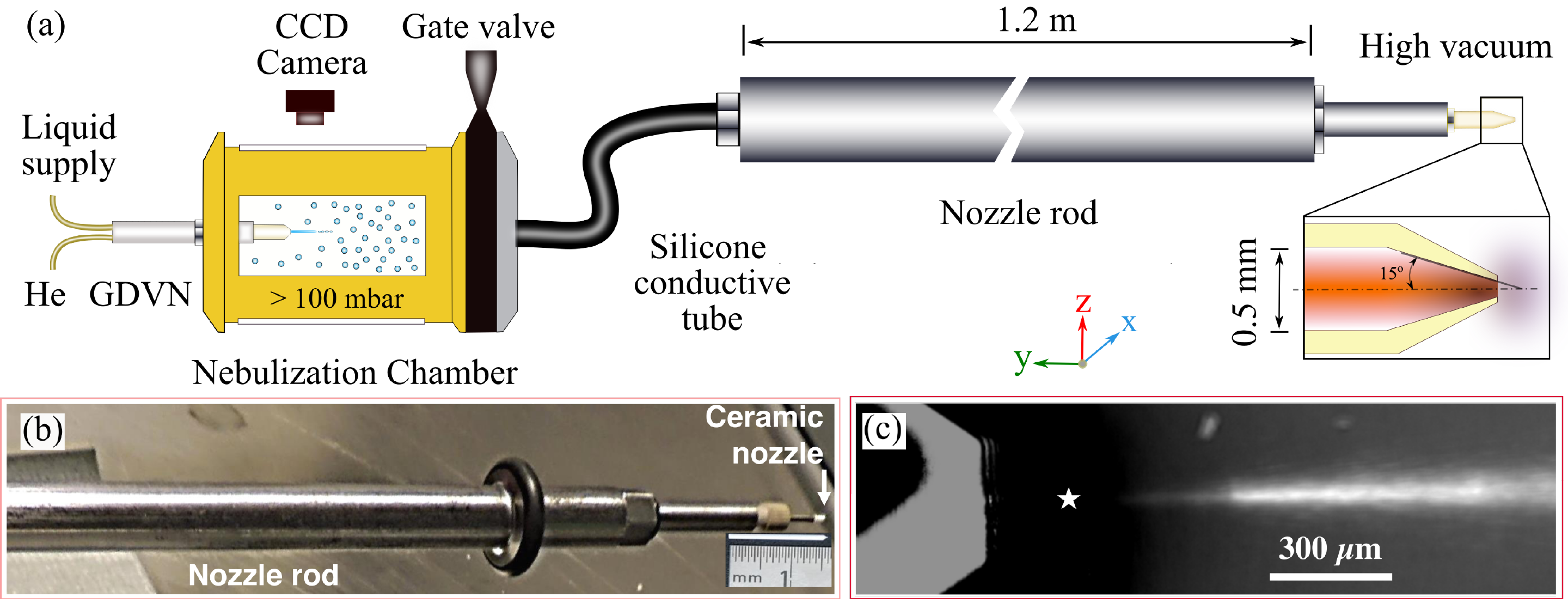}
		\caption{CNAI assembly and its operation during the CXI experiment. (a) Sketch of the basic
			aerosol generation and transportation setup. (b) The aerosol nozzle mounted on the nozzle rod.
			(c) Time integrated image of a laser-illuminated stream of GV particles exiting the CNAI,
			recorded using the in-line microscope at the CXI instrument. This image was formed by
			averaging over 3.7~min, with a running median background subtracted from each frame. The CNAI
			tip is seen in the left portion of the image, and the approximate x-ray focal point is
			indicated by the star.}
		\label{fig:jet}
	\end{figure}

	The granulovirus suspension was aerosolised using a gas dynamic virtual nozzle
	(GDVN)~\cite{Beyerlein:RSI86:125104} mounted in a cylindrical nebulization chamber as depicted in
	\autoref{fig:jet}~(a). A GDVN uses gas flow focusing to create a liquid jet with a diameter
	significantly smaller than the orifice of the nozzle, and which consequently breaks up to form a
	mist of droplets. The liquid was pressurized to flow from the nozzle at rates between
	$\unit{2.7}{\micro}$l/min and $\unit{3.5}{\micro}$l/min, producing droplets of about
	$\unit{2}{\micro\metre}$ diameter at a rate between $11\times10^6\text{~s}^{-1}$ and
	$14\times10^6\text{~s}^{-1}$, each containing on average 1.3 nanocrystals. This is equivalent to
	particle flow rates of $8.6\times10^8$--$1.1\times10^9$ particles/min.
	
	The focusing gas was helium, which was set to a mass flow rate in the range of 10--60~mg/min. The
	nebulisation chamber had an inner diameter of approximately 40~mm and was 150~mm in length, giving a
	residence time in the chamber of several minutes, and a helium pressure that stabilised at a value
	between 100 mbar to 1 bar. Under
	these conditions most of the solvent evaporated to produce nanocrystals suspended in a humid helium
	atmosphere~\cite{Kirian:SD2:041717}. Drops that contained more than one particle during the initial
	stage most likely formed clusters of crystals \cite{Cho:Langmuir23:12079,Daurer:IUCrJ4:3}. 
	The aerosol flowed through conductive silicone rubber tubing (Simolex,
	6.3~mm inner diameter, 30~cm length), which was coupled to a standard ``nozzle rod'' of the CXI
	beamline, which is a 1.2~m long stainless steel tube with a 6.3~mm inner
	diameter that is normally used to transfer liquid-jet injectors in and out of the main
	experimental chamber without breaking vacuum~\cite{Weierstall:RSI83:035108}. The conductive tubing
	along the entire particle path acted as a Faraday cage to shield external electric fields from
	interacting with particles that might become charged through triboelectric effects in the GDVN. The
	aerosol finally exited the CNAI, which was mounted at the end of the CXI nozzle rod much like a
	typical liquid-jet nozzle. It consisted of a ceramic injection-molded tube of 1~mm outer diameter,
	$\unit{500}{\micro\meter}$ inner diameter, a short converging section with a convergence angle of 15$^{\circ}$, and a
	$\unit{100}{\micro\meter}$ exit aperture (further details can be found in our previous work \cite{Kirian:SD2:041717}).
	%Within the main chamber, the CNAI was furthermore contained within a differentially pumped shroud
	%that helps maintain low pressure in the main chamber. 
	%The position of the entire nozzle rod, and
	%hence also the CNAI, could be adjusted with the aim of placing the focus of the nanocrystal beam
	%from the CNAI coincident with the x-ray focus.
	
	During the diffraction experiment we monitored the crystal injection through direct optical imaging
	of scattered laser light from injected particles~\cite{Awel:OE24:6507}. A pulsed Nd:YLF laser
	(527~nm, $\sim$3~mJ per $\sim$150~ns pulse, 120~Hz) was focused to a $\sim$0.8~mm spot within the
	aerosol stream, and scattered light was observed through the in-line microscope available at CXI
	(Questar long distance microscope, model: QM-1 MK III, NA~=~0.05 at 750 mm objective distance).
	Images were recorded using an OPAL-4000 CCD camera and stored at 30~Hz. \autoref{fig:jet} (c) shows
	a 3.7~min time-averaged optical image of particles exiting the injector. We determined that
	particles moved at speeds of approximately 300~m/s when they exited the injector, to arrive at the
	x-ray interaction point within a flight-time of less than $\unit{1}{\micro\second}$. This particle
	speed was evaluated from the streak length of recorded particle images produced by laser
	illumination with a known pulse duration, conducted during laboratory characterization of the CNAI
	(see \autoref{fig:GV}~(a) and \autoref{sec:characterization}). The CNAI tip is seen to the left of
	\autoref{fig:jet} (c), and the approximate x-ray focal point is indicated by the star. The particle
	stream could not be observed at points close to the CNAI tip because direct scattering from the tip
	saturated the imaging CCD.
	
	\section{Injector characterization and hit-fraction estimates}
	\label{sec:characterization}
	In order to develop and characterize the operation of the CNAIs we conducted tests of both
	15$^{\circ}$ and 30$^{\circ}$ CNAIs in our laboratory. The setup differed from our previous
	work~\cite{Kirian:SD2:041717} by the inclusion of a narrow particle transport tube intended to
	replicate the delivery system used at CXI. Aerosolized GV particles were transported from the
	nebulization chamber to the CNAI tip using stainless steel tubing of 4~mm inner diameter and 700~mm
	length. The GDVN was operated at flow rates of $\unit{2.7}{\micro}$l/min and 28~mg/min for liquid
	sample and helium, respectively. A GV concentration of approximately $1.6\times10^{9}$ particles/ml
	was used (this was diluted by a factor of 200 from the solution that was used at the CXI
	experiment). This flow rate and sample concentration corresponds to the generation of drops at a
	rate of approximately $1.1\times10^7\text{~s}^{-1}$ and an entrance rate of aerosolized particles of
	$7.2\times10^4\text{~s}^{-1}$.
	
\begin{figure}
		\centering
		\includegraphics[width=\linewidth]{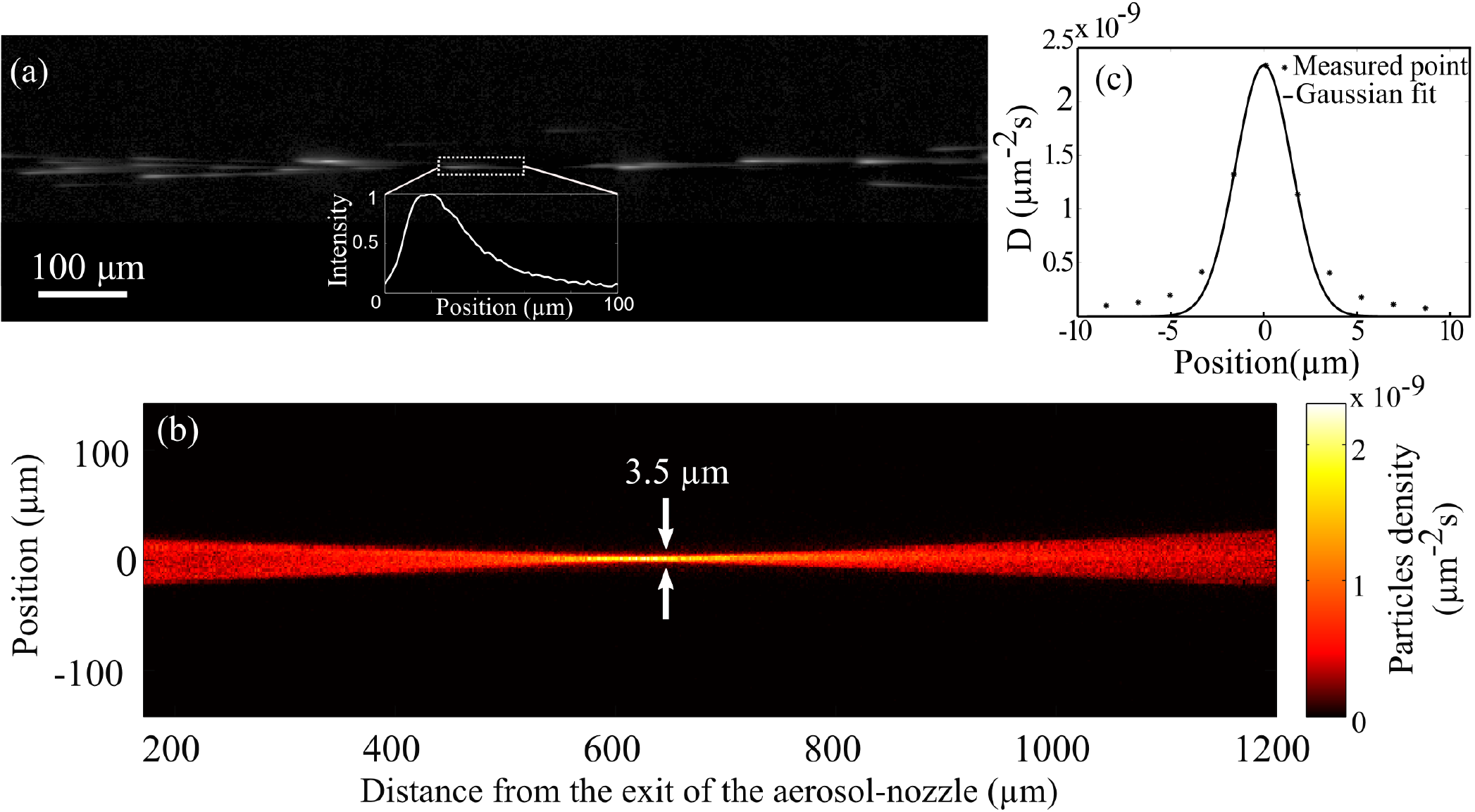}
		\caption{Laboratory characterization of a beam of GV particles focused with the 15\,$^{\circ}$
			convergent aerosol-nozzle using a strong-magnification imaging microscope. (a) A single
			exposure showing streaked images of GV particles caused by the 100 ns laser illumination. The
			particles are moving from left to right and their streaked images have nonuniform intensity
			due to the relatively slow decay of the illumination laser pulses. (b) The two-dimensional
			rate-corrected particle density determined from the centroids of individual particle images
			such as the one shown in (a). (c) Gaussian fit to the particle density at the focal plane in
			(b).}
		\label{fig:GV}
	\end{figure}

	The imaging setup used for visualizing particles was described in detail
	previously~\cite{Awel:OE24:6507}. Briefly, it was comprised of a Nd:YLF laser (Spectra Physics
	Empower ICSHG-30, 527~nm, approximate pulse duration 100~ns, repetition rate 1~kHz, pulse energy
	20~mJ) to illuminate particles, a high-frame-rate CMOS camera (Photron SA4) and a 5$\times$ magnification, 0.14 NA
	microscope objective to record images. The laser beam was collimated to a 2~mm spot, such that it
	illuminated particles across the entire field of view of the camera. The camera exposure time was
	set to 20~ms, such that each frame contained 20 pulses of the 1~kHz Nd:YLF laser illumination. A
	single image of particles emerging from the CNAI is shown in \autoref{fig:GV}~(a). The images are
	streaked due to the high velocity of the particles, and the observed intensity profile of these
	streaks reflects the relatively fast rise and slow decay of the Nd:YLF laser pulse. Centroid
	positions of individual particle streaks contained in 23,500 frames were used to produce the
	rate-corrected two-dimensional particle density map shown in \autoref{fig:GV}~(b). This
	rate-corrected density has units of particles per area \emph{per particle generation rate} and is
	defined as
	\begin{equation} D =
		\frac{N_p}{A\times{}R} \; , \label{RCD}
	\end{equation}
	where $N_p$ is the average number of particles that fall within a spatial bin of area $A$, and $R$
	is the rate at which particles entered the injector. Note that $N_p$ represents the average particle
	counts at an instant in time and not a time-integration over many exposures, which is appropriate
	because we intend to use the particle injector with femtosecond pulses. In our case, $N_p$ was
	computed by summing the number of particles that fell within each spatial bin, and then dividing by
	the number of recorded images and the number of laser illumination pulses per image.
	
	The measured rate-corrected density $D$ may be used to estimate the optimal hit fraction that could
	be achieved under idealized conditions in our x-ray measurements. If an entrance rate of $R_X$ is
	used in the x-ray measurements, the 2D particle number density is $D R_X$. We define the effective
	cross-sectional area $\sigma$ such that the average number of particles intercepted by an x-ray
	pulse is $\sigma D R_X$. Assuming Poisson statistics, the probability of intercepting just one
	particle in an x-ray pulse is
	\begin{equation}
		H_1 = \exp(-\sigma D R_X) \sigma D R_X \approx \sigma D R_X
	\end{equation}
	where the approximation holds to within $\sim$10\% error as long as $ \sigma D R_X < 0.1$. We define
	the x-ray beam diameter as $d_X$ and the particle beam diameter as $d_p$ and estimate two limiting
	cases for the effective cross sectional area. The first case, $\sigma^+=\frac{\pi}{4}(d_x+d_p)^2$,
	describes the optimistic limit in which a particle at the periphery of the x-ray beam produces
	acceptable diffraction. The second case, $\sigma^-=\frac{\pi}{4}(d_x-d_p)^2$, corresponds to the
	stronger assertion that an acceptable diffraction pattern requires that the entire x-ray-beam width
	falls within the particle (if $d_X<d_p$) or that the entire particle falls within the x-ray-beam
	width (if $d_p<d_X$). Finally, we arrive at two limiting hit-fraction estimates:
	\begin{equation}
		H_1^\pm \approx \frac{\pi}{4} D (d_X \pm d_p)^2 R_X
	\end{equation}
	
	The maximum rate-corrected particle density recorded in the laboratory, i.~e., at the focus of the
	particle beam shown in \autoref{fig:GV} (b), was
	$D\approx{\unit{2.2\times10^{-9}}{\micro\meter}}^{-2}\text{~s}$. Assuming the approximate values
	$d_X\approx150$~nm, $d_p\approx300$~nm, and $R_X\approx11\times10^6\text{~s}^{-1}$ suggests that the
	maximum hit fraction to be expected in our XFEL diffraction measurements is in the range
	$H_1^-\approx 0.04$~\% to $H_1^+\approx0.4$~\%.	This predicted hit fraction is much higher than the hit fraction we achieved during the CXI experiment, as discussed in the next section.	
	
	\section{X-ray diffraction analysis and discussion}
	\label{sec:diffraction}
	Diffraction measurements were conducted at a photon energy of 8~keV and an estimated average pulse
	energy of 4.2~mJ prior to the $\sim$30-50~\% beamline transmission
	losses~\cite{Boutet:2016:pricom}. The CSPAD detector was located 127.9~mm downstream from the x-ray
	focus. We recorded detector data frames for every X-ray pulse, at a rate of \unit{120}{\hertz}, for a
	cumulative total of 1.3 hours, which resulted in approximately 560,000 data frames.
	
	In all of our diffraction analysis we excluded all pixels from each detector frame that had abnormally high or low variances
	or mean values in ``dark'' measurements made without x-rays, as well as a few patches of pixels for
	which there was obvious stray-light background. For every frame, the dark measurement was
	subtracted, and then a uniform common-mode electronic noise constant was subtracted from each
	detector panel. The common-mode offset was determined from unbonded detector pixels that are not
	sensitive to x-rays. The detector gain relating detector digital units to photon counts per pixel
	was obtained from a histogram of the pixel values, which yielded clear peaks corresponding to counts
	of zero, one, and two photons. Most of this analysis was performed using the
	Python \texttt{psana} package provided by LCLS \cite{Damiani:IUCrJ49:672}.
	
\begin{figure}
		\centering
		\includegraphics[width=\linewidth]{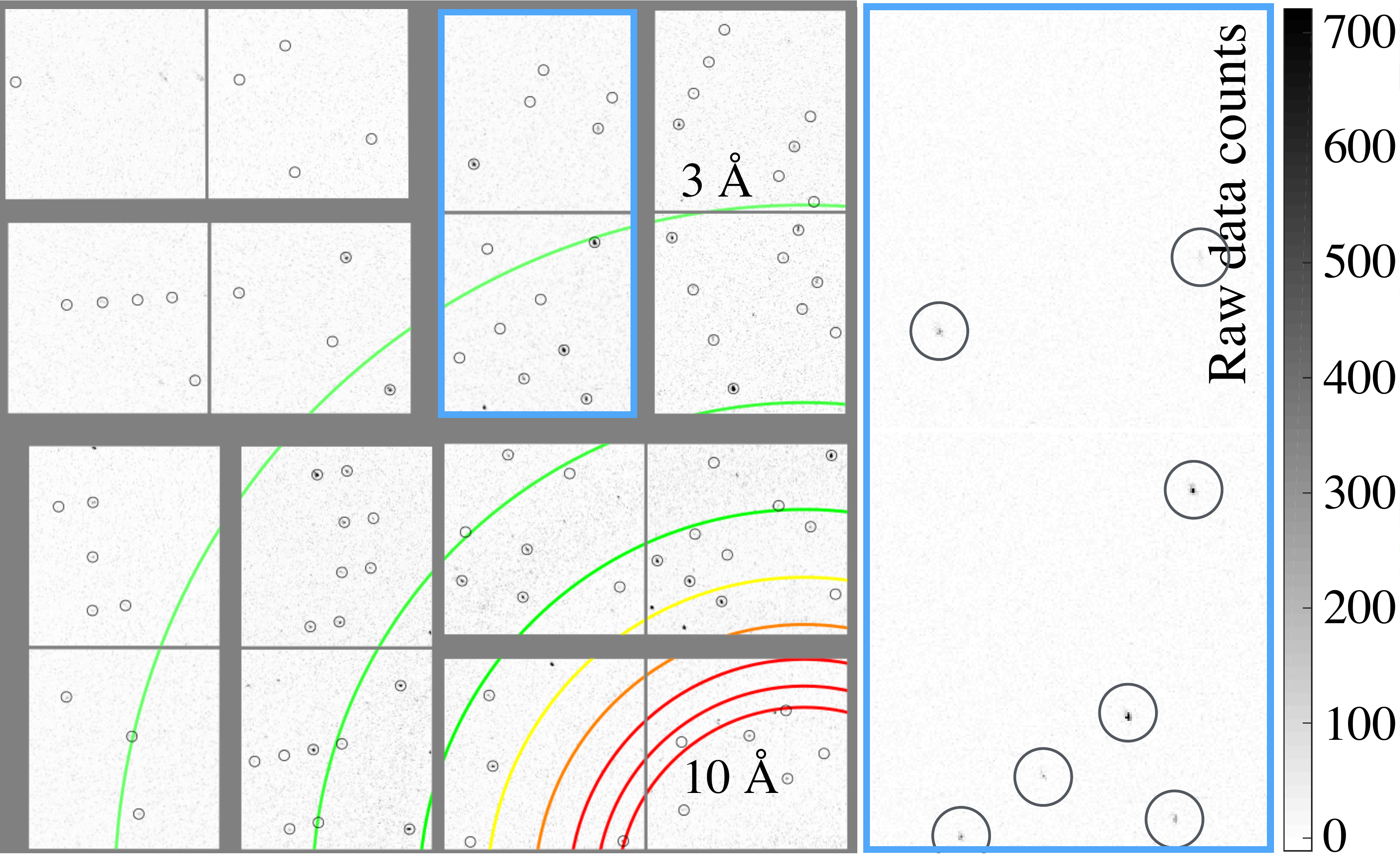}
		\caption{One detector quadrant of an indexed diffraction pattern obtained from aerosolized GV
			crystals. The colored rings indicate the resolution from 10~\AA~to~3~\AA, in steps of 1~\AA.
			The gray circles in the left-hand panel indicate the expected locations of Bragg peaks as
			determined by auto-indexing in the CrystFEL software suite~\cite{White:JACR45:335}. The
			right-hand panel shows expanded view of an individual detector tile, marked by the blue
			rectangle on the left. Circles in this expanded-view panel indicate peaks that are easily
			recognizable by eye. Notably, the predicted peak locations indicated by CrystFEL do not
			perfectly agree with those that the human eye notices, but this is typical of first indexing
			results and could be improved through the CrystFEL post-processing routines.}
		\label{fig:gv-aerosol}
	\end{figure}

	\autoref{fig:gv-aerosol} shows one quadrant of a recorded diffraction pattern from an aerosolized GV
	crystal, where the average detector dark frame and common-mode offsets have been subtracted. A total
	of 33 hits from GV were recorded, corresponding to a hit fraction of $\sim\!0.006$~\%. 24 patterns
	(73~\% of hits) were indexed using the CrystFEL software suite~\cite{White:JACR45:335}. Autoindexing
	failed on patterns that appeared to consist of multiple crystals clumped together. We expect the hit
	fraction for our aerosol injector to be significantly lower than a typical liquid jet (about 1-10\%)
	because of the $\sim$25-fold higher particle speed of the aerosol beam and the $\sim$4-fold
	reduction in liquid flow rate. However, our recorded hit fraction was still lower than the range
	$0.04 - 0.4$~\% that we estimated from our laboratory measurements.
	
	For comparing the background obtained using the CNAI to that typically observed in liquid-jet
	experiments we examined data from a previous SFX experiment~\cite{Oberthuer:SR7:44628} in which the
	exact same GV sample was injected into the x-ray beam as a liquid suspension with a GDVN. All
	experimental parameters were identical in both the CNAI and GDVN measurements except for the pulse
	energy, which was 4.6~mJ on average for the GDVN measurements.
	
\begin{figure}
		\centering%
		\includegraphics[width=\linewidth]{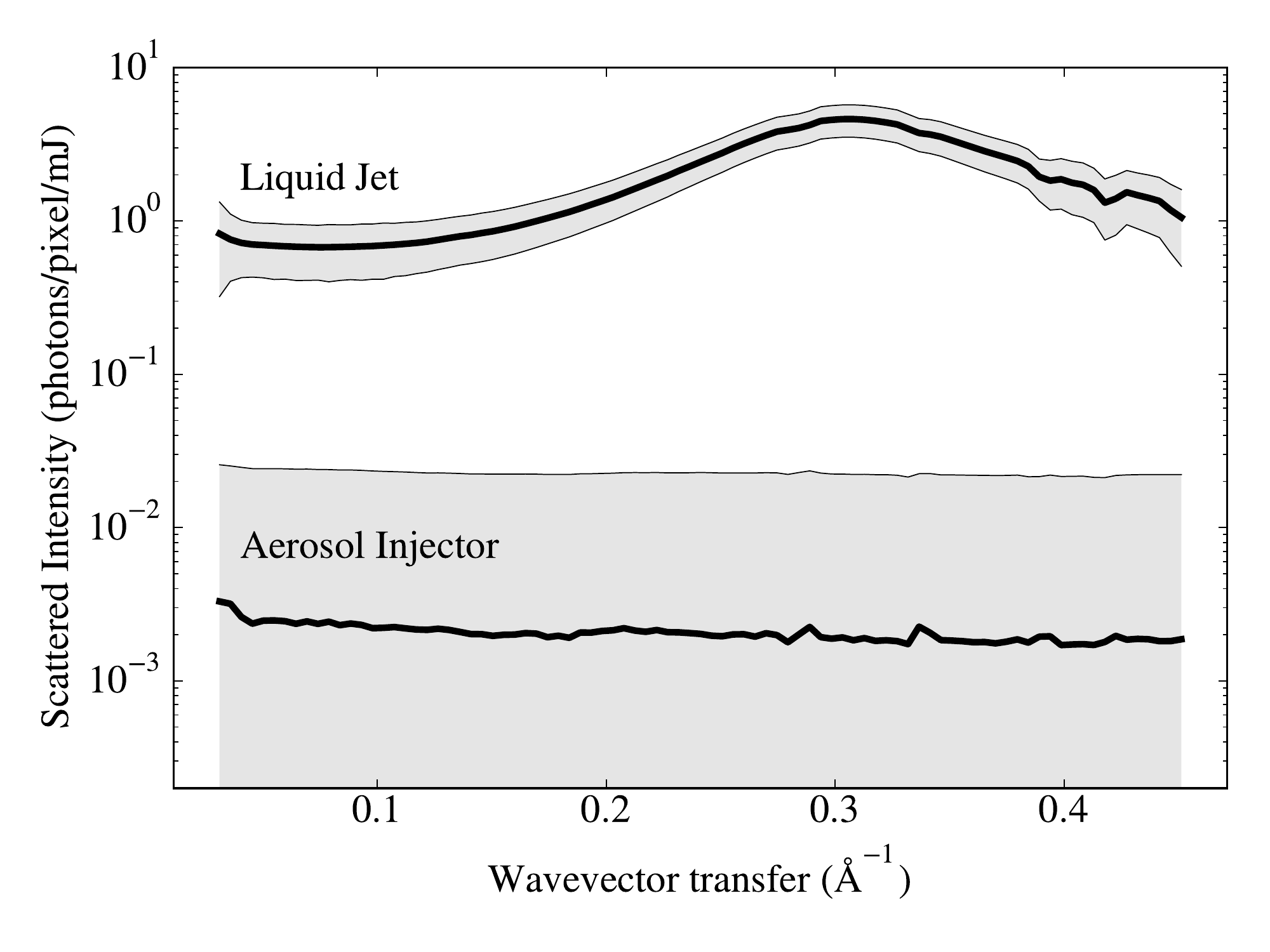}
		\caption{% \old{Histogram of 4,500 background radial
			Average radial intensity profiles, on a logarithmic scale, for data measured using the GDVN
			(labeled ``Liquid Jet'') and the CNAI (labeled ``Aerosol Injector'') injectors. The average
			per-pixel standard deviations determined from more than 10,000 frames are indicated by the
			vertical width of the gray regions. After averaging, the profiles and standard deviations were
			normalized by dividing by the average pulse energy, and then divided by the digital-to-photon
			conversion factor of 18.3. The horizontal axis corresponds to the wavevector transfer
			$q=2\sin(\theta)/\lambda$ where $\theta$ is the Bragg angle and $\lambda$ is the wavelength.}
		\label{fig:background}
	\end{figure}

	The comparison of background scattering for the CNAI and GDVN approaches is presented in
	\autoref{fig:background}, which shows a plot of the normalized azimuthally-averaged profiles of
	scattered-photon counts (per pixel and per mJ of pulse energy), as a function of photon-wavevector
	transfer. The per-pixel standard deviations in the measurements are indicated by the gray regions in
	\autoref{fig:background}. The average profiles were divided by the average pulse energy to account
	for the slightly higher pulse energy in the case of the GDVN. The frames used in
	\autoref{fig:background} were sampled uniformly from the final $\sim$5 minutes of data collection,
	when the conditions were closest to optimal, although little difference was noticed in other
	measurement segments. We excluded frames that fell below 1~mJ pulse energy. We
	additionally excluded frames that were visually corrupt as well as those for which the x-rays
	obviously missed the liquid jet, which corresponded to less than 10\% of the frames. After removing
	these outliers, we confirmed that more than 10,000 frames contributed to each of the two profiles.
	
	As can be seen from the plots in \autoref{fig:background}, the liquid jet produces a background that
	is over 1000 times higher at a wavevector transfer of $q = 2 \sin(\theta) / \lambda = 0.32 \text{ \AA}^{-1}$, corresponding to a resolution of 3.1~\AA, where $\theta$ is the Bragg angle and $\lambda$ the
	wavelength. This coincides with the mean distance between oxygen atoms in water where diffuse
	scattering from water has its maximum. At low scattering angles the background from the liquid jet
	was about 200 times higher than for aerosol injection. The liquid jet for these measurements was
	operating at a flow rate of $\unit{20}{\micro}$l/min. Typical liquid flow rates needed to produce a
	stable jet range from $\unit{5-30}{\micro}$l/min, depending on the viscosity and surface tension of
	the liquid and the nozzle geometry. The volume of liquid that interacts with the x-ray beam scales
	roughly as the square root of the volumetric flow rate~\cite{Beyerlein:RSI86:125104}, and thus the
	liquid jet background is rather typical.
	
	Due to our convergent micro-focused particle beam, hit fractions are highly sensitive to the
	relative positioning of the CNAI with respect to the x-ray beam. Our initial diagnostic for particle
	beam positioning was direct imaging of scattered light, which allowed for the rough positioning of
	the CNAI. From this initial position, it was necessary to perform a subsequent two-dimensional scan
	of the injector position in an effort to optimize the spatial overlap between particle beam focus
	and x-rays. Due to the limitations of our 6-hour measurement shift, we only performed one $\unit{200}{\micro\meter}$ scan in the direction transverse to the particle beam and one $\unit{400}{\micro\meter}$ scan along the particle beam direction. It is therefore highly unlikely that we located the ideal position that maximizes the hit fraction. However, we expect that the background scatter we observed is representative of the gas
	and water vapor exiting the injector because the gas expansion into vacuum is highly divergent.
	Direct imaging of the gas density leaving the CNAI~\cite{Horke:JAP121:123106} shows that the gas
	plume spans volume hundreds of micrometers wide around the XFEL beam position.
	
	Another possible culprit for our sub-optimal hit fraction is a sub-optimal aerosol transmission
	efficiency, which might be remedied by reducing the overall transportation tube length, increasing
	the particle generation rate, decreasing the particle speed, increasing the volumetric flow rate of
	carrier gas, or by the addition of aerodynamic lenses within the transport tube, which would
	maintain particles near the center of the transport tube. Although aerosol injection hit fractions
	tend to be relatively low in comparison to liquid jets, recent work at at the CXI instrument
	reported hit fractions of $0.83~\%$ for aerosolized 40 nm viruses delivered with an aerodynamic lens
	stack aerosol injector~\cite{Daurer:IUCrJ4:3}.

	Although it is convenient that our miniaturized CNAI is compatible with standard GDVN mounting
	hardware, the downside is that the small exit aperture, $\unit{100}{\micro\meter}$ diameter in our
	case, is prone to clogging. We have successfully operated our CNAIs in the laboratory for many hours
	without interruption, but clogging typically occurs whenever the aerosolization liquid jet
	misbehaves and produces large droplets for a period of a few minutes. It is, therefore, essential to
	ensure the formation of small droplets and continuous flow of carrier gas. In the XFEL experiment
	reported here, there were a total of three clogged aerosol-nozzles, each of which required
	$\sim$20-30 minutes to replace. The severity of this issue could be greatly reduced by filtering out
	large droplets with, for example, an inline impactor \cite{Maenhaut:NIM109:482}, and by using
	electrospray ionization to produce smaller initial droplet diameters \cite{Yamashita:JPC88:4451,Chen:JAS26:963,Bogan:NanoLett8:310}.

	It must finally be noted that the GV crystals utilized here are notoriously robust and survive in
	nearly pure water. For crystals that dissolve, for instance, upon varying pH, it may be feasible to
	avoid droplet evaporation by using a humidified carrier gas, by using electrospray nebulization, or
	by simply placing the nebulization source close to the entrance of the aerosol nozzle to reduce the
	time of transport.

	\section{Conclusions}
	\label{sec:conclusion}
	We demonstrated x-ray diffraction from aerosolized sub-micrometer protein crystals with background
	levels drastically lower than in typical SFX experiments utilizing liquid jets. This may be
	important for coherent-diffractive-imaging experiments on weakly scattering targets such as isolated
	proteins, viruses, or cells, as well as for the measurement of diffuse scattering or
	lattice-transform signals between crystalline Bragg
	reflections~\cite{Ayyer:Nature530:202,Kirian:PRX5:011015}. We showed that our injector is compatible
	with the existing hardware at LCLS, allowing quick changes from a liquid jet to an aerosol injection
	system in a single experiment. The relatively high ($\sim$300~m/s) particle velocities may be useful
	for avoiding damage due to x-ray induced explosions when using new XFEL sources with pulse
	repetition rates up to 4.5~MHz.
	
	While the obtained 0.006\% hit fraction at LCLS was much lower than in typical liquid jet x-ray
	diffraction experiments, laboratory measurements suggest that this can be improved by orders of
	magnitude. Based on these laboratory measurements, we suspect that the low hit fractions observed in
	this study are a result of aerosol transport losses, clustering of particles, clogging of the
	aerosol-nozzle due to an under-performing GDVN nebuliser, or misalignment between the x-ray focus
	and particle beam focus. As we have noted, there are several possible routes to improve upon the
	injection strategy described here, as evidenced by other aerosol injection work performed at the
	same CXI instrument~\cite{Daurer:IUCrJ4:3}.
	
	Above all, the lower background achieved with the aerosol nozzle somewhat offsets the lower hit
	fraction, since the number of required measurements depends inversely on the square of the signal to
	noise ratio of intensities, or directly proportional to the background counts.
	
	This proof of principle experiment was performed on granulovirus occlusion bodies suspended in
	water. These protein crystals have naturally evolved to be robust against the change in the buffer
	conditions and dehydration caused by evaporation of the liquid layer on the crystals surface.
	However, most protein crystals are not stable in pure water. When working with other types of
	crystals, the liquid buffer evaporation rate on the surface of the crystals must be controlled, for
	example by controlling the relative humidity at the crystals
	\cite{Sanchez-Weatherby:ACD65:1237,Roedig:SR5:10451}.

	\ack{\emph{\textbf{Acknowledgments}}}\\
	We thank the LCLS staff for accommodating laser-light-scattering imaging in the CXI instrument
	during our experiments. Portions of this research were carried out at the LCLS at the SLAC National
	Accelerator Laboratory. This LCLS beamtime was part of the Protein Crystal Screening (PCS) program.
	LCLS is an Office of Science User Facility operated for the US Department of Energy Office of
	Science by Stanford University. Use of the Linac Coherent Light Source (LCLS), SLAC National
	Accelerator Laboratory, is supported by the U.~S.\ Department of Energy, Office of Science, Office
	of Basic Energy Sciences under Contract No. DE-AC02-76SF00515. Parts of the sample delivery system
	used at LCLS for this research was funded by the NIH grant P41GM103393, formerly P41RR001209.
	
	In addition to DESY, this work has been supported by the excellence cluster ``The Hamburg Center for
	Ultrafast Imaging--Structure, Dynamics and Control of Matter at the Atomic Scale'' of the Deutsche
	Forschungsgemeinschaft (CUI, DFG-EXC1074), the Gottfried Wilhelm Leibniz Program of the DFG, the
	European Research Council under the European Union's Seventh Framework Programme (FP7/2007-2013)
	through the Synergy Grant AXSIS (ERC-2013-SyG 609920) and the Consolidator Grant COMOTION
	(ERC-Küpper-614507), the Helmholtz Association ``Initiative and Networking Fund'', and the
	Australian Research Council's Discovery Projects funding scheme (DP110100975). R.A.K.\ acknowledges
	support from an NSF STC Award (1231306).

	% figure 1
	
	% Figure 2
	
	% figure 3
	
	\clearpage
	\referencelist[aerosol-sfx]
\end{document}